\documentclass[11pt,a4paper]{article}
\usepackage[hyperref]{eacl2021}

\usepackage{amsmath}
\usepackage{times}
\usepackage{url}
\usepackage{amssymb}
\usepackage{xcolor}
\usepackage{latexsym}
\usepackage{subfig}
\usepackage{varwidth}
\usepackage{graphicx}
\usepackage{cases}
\usepackage{placeins}
\usepackage{array, boldline, makecell, booktabs}
\usepackage{multirow}
\usepackage{multicol}
\usepackage{latexsym}
\usepackage{graphicx}
\usepackage{placeins}
\usepackage{booktabs}

\newcommand{\comment}[1]{}
\makeatletter
\def\thickhline{%
  \noalign{\ifnum0=`}\fi\hrule \@height \thickarrayrulewidth \futurelet
   \reserved@a\@xthickhline}
\def\@xthickhline{\ifx\reserved@a\thickhline
               \vskip\doublerulesep
               \vskip-\thickarrayrulewidth
             \fi
      \ifnum0=`{\fi}}
\makeatother
\usepackage{xcolor,colortbl}
\newlength{\thickarrayrulewidth}
\definecolor{Gray}{gray}{0.85}
\usepackage{subfig}
\usepackage{bm}

\usepackage{microtype}

\aclfinalcopy 
\def\aclpaperid{1342} 
\definecolor{blue_c}{RGB}{40, 116, 166}
\definecolor{orange_c}{RGB}{175, 96, 26 }
\newcommand\BibTeX{B\textsc{ib}\TeX}
\newcolumntype{?}{!{\vrule width 3\arrayrulewidth}}

\title{Correlated Feature Selection for Tweet Spam Classification}

\author{Prakamya Mishra \\
  Shiv Nadar University\\
  \texttt{pkms.research@gmail.com} \\}

\date{}

\begin{document}
\maketitle
\begin{abstract}
The identification of spam messages on social networks is a very challenging task. Social media sites like Twitter \& Facebook attracts a lot of users and companies to advertise and attract users of personal gains. These advertisements most of the time leads to spamming, which in return leads to poor user experience. The purpose of this paper is to undertake the analysis of spamming on Twitter. To classify spams efficiently, it is necessary to first understand the features of the spam tweets as well as identify attributes of the spammer. We extract both tweet based features and user-based features for our analysis and observe the correlation between these features. This step is necessary as we can reduce the training time if we combine the highly correlated features. Our proposed approach uses a classification model based on artificial neural networks to classify the tweets as spam or non-spam giving the highest accuracy of 97.57\% when compared with four other standard classifiers namely, SVM, K Nearest Neighbours, Naive Bayes, and Random Forest.
\end{abstract}

\section{Introduction}
Online social networking platforms such as Twitter, Facebook, Instagram, etc. allow people to meet, discuss, and work together by collaborating on projects with just a click. The combined number of users just on Facebook, Linkedin, Instagram, and Twitter stands at 3200 million as of September 2017 \footnote{https://www.statista.com/statistics/282087/number-of-monthly-active-twitter-users/}. Twitter has generated a lot of interest among netizens recently due to its widespread use by influential people like the  Presidents and the Prime Ministers of powerful countries.  As per the latest reports, approximately 330 million active users are on Twitter .

One of the interesting properties of Twitter is the ability to follow any other user with a public profile. Media organizations, politicians, and celebrities are reaching millions of followers every day. It is interesting to note that in many cases, the number of actual followers is not genuine \cite{Ghosh:2012:UCL:2187836.2187846}. In a recent incidence in India, some of the leading national newspapers published headline news on Oct 21, 2017, with the title being "\textit{Bots behind the rise in XXXX Twitter popularity?}" when one of his tweets received 30,000 re-tweets.

Fake Twitter followers and Robot driven accounts being used for re-tweets are not new phenomena. There are several softwares in the market such as Twitter bots that use Twitter APIs to control Twitter accounts. These software bots can be used to send tweets, re-tweets, follow, unfollow, and increase the number of likes on a tweet \cite{Thomas:2011:SAR:2068816.2068840}. Varol et al. \cite{DBLP:journals/corr/VarolFDMF17} in his paper states that as many as 48 million accounts on Twitter are actually bots. This means that approximately 15\% of the profiles are fake. The growth in popularity of Twitter in recent years has led to a large number of spammers who exploit and manipulate these numbers to misuse the whole medium for unwanted gains. Spamming is not a new concept. Initially, the term was associated with bulk email which was unsolicited. Lots of research has been done to overcome this problem and now we have fairly accurate filters that keep on segregating spam mails into a separate folder. 

The internet society in 2015 estimated that 85\% of the global emails are spam \footnote{https://www.internetsociety.org/policybriefs/spam}. Considering that decades of research has been done in the area of email filtering, Tweet spam filtering needs significant contribution from the academic research community. Twitter defines spam as unsolicited, repeated actions that negatively impact other people. Examples of the aforementioned could be, posting harmful links of phishing sites, using automated bots for mass following, abusing, creating multiple handles, posting, or re-tweeting on trending topics unnecessarily.

Twitter has its own mechanism of spam detection however it is still in its infancy and the academic research community should provide its support. On average Twitter is able to detect roughly 3.2 million suspicious accounts per week \footnote{http://www.dailymail.co.uk/sciencetech/article-4931430/Twitter-reveals-closes-3-2m-spam-accounts-WEEK.html}. The primary issue that needs attention is to understand how harmful these spams can be. To figure out a solution to this problem, we note that the US intelligence community released a report in January 2017 highlighting the role that Russia Today (RT) might have played in influence the 2016 U.S. Elections. This is a big allegation and which is still under investigation; however, it still shares the influence of online platforms particularly Twitter for such big events \cite{FM7090} .

In section 2, we will review related spam detection research. In section 3, we will explain our process of data collection and preprocessing. In section 4, we will discuss our proposed approach for feature selection which is further divided into 2 sub-sections for the analysis of Tweet-based features and User-based features. In section 5, we detail our experiments and comparisons with our spam detection model. Section 6 concludes this research with directions to future work.

\section{Related Work}
Twitter continues to gain popularity among the various social networks currently available and thus attracts spammers who would try to abuse the system by manipulating existing features to gain undue advantage \cite{Thomas:2011:SAR:2068816.2068840}. There have been numerous studies that have proposed machine learning and artificial intelligence techniques for detecting spammers. Existing studies have focussed on classification algorithms to distinguish between spammers and non-spammers \cite{Ghosh:2012:UCL:2187836.2187846}. Lee et al. \cite{Lee:2010:USS:1835449.1835522} created social honeypots for the identification of spammers. Many studies have focused on URL based spam detection and blacklists based on domain and IP address. This has not been successful since short URL's obscure the base and new short URL's are used by spammers as soon as old ones are blacklisted \cite{6679991}. Grier et al. worked extensively on blacklisted URL's \cite{Grier:2010:SUC:1866307.1866311}.

Twitter identifies users through a unique username referred to as the screen name. Each user can send replies containing screen names. One can also mention another user's screen name anywhere in their tweet. This feature helps users to track conversations and know each other. Spammers, however, use this feature by including many screen names in their replies and tweets, If there are too many replies or mentions in tweets by a user, Twitter will treat this as suspicious\footnote{Twitter support https://help.twitter.com/} \cite{Stringhini:2010:DSS:1920261.1920263}.

Twitter allows a message with a maximum length of 140 characters. Due to this restriction, many URLs are shortened in the tweets. However, short URLs can obscure the source and this property has been used by spammers to camouflage the spam URLs \cite{6679991}. Twitter allows the re-tweets and all such re-tweets start with \textit{@ RT}. Many authors  \cite{kabakus2017survey} use the number of re-tweets in the most recent 20-100 tweets of a user as an important feature in spam detection.\\
Trending topics has become ubiquitous topic these days. If there are several tweets with the same term, then it will become a trending topic. Spammers seek attention by posting many unrelated tweets with trending terms \cite{5741690}. Another prominent feature of Twitter allows users to create a public and private list to categorize people in different groups based on similar interests \cite{kim2010analysis}.

There are two major categories in which we can segregate the extracted features for Twitter spam detection. User profile based features and tweet content-based features. Some of the major user profile based features are the number of followers, the number of follows, duration of the existing account, the number of favorites, number of lists in which the user has membership, and the average number of tweets a particular user sends \cite{kabakus2017survey}. The tweet content-based features are the number of times a particular tweet has been re-tweeted, the number of hashtags, the number of times a particular tweet has been mentioned, the number of URLs included in a tweet, the number of characters and number of likes in this tweet \cite{kim2010analysis}. 

\section{Data collection}
We need a labeled dataset to train and test our model. A majority of spam contains embedded URLs and many researchers have focused only on this \cite{5958045}. For making our labeled dataset of spam and non-spam tweets, we have used tweets from spammers and non-spammers. We have used the list of spammers from the reference \cite{benevenuto@ceas10} and for non-spammers, we have randomly picked Twitter user accounts. For each user, we have extracted at most 100 previous tweets using Twitter API Tweepy by giving the user screen name as a query. Along with the text of the tweets, we have also extracted the timing of the tweets, the number of previous tweets, favorites, friends, lists, number of followers of the users.\\
We have extracted 719300 tweets from an of total 760 users out of which 370 (350900 tweets) users are randomly picked non-spammers and 390 (368400 tweets) are spammers.

\section{Feature selection}
Any machine learning based spam classification technique would need feature extraction. Historical information of user-based features such as the number of tweets sent by the user in the last 30 days etc. are important and give useful insights. To make sure that feature extraction is a real-time process, we have used lightweight features from Tweepy API and derived new features from these extracted features.

One of the important differences between spammers and non-spammers is the intent of spamming. Spammer's purpose of tweeting is to get some undue advantage through those tweets or belittle some rivals. Taking into account the intentions of the tweets it is obvious that the spammer's tweets should have distinct characteristics compared to non-spammer tweets. Historical data suggests that the average time spent by non-spammers should be less than spammers \cite{FM2793}. There can be various other discriminatory attributes that can reflect on user behaviors.

In this section, we will take into account the features that have been considered for each tweet for classification. Each tweet has two broad categories of features viz. tweet based features, such as those which are related to that particular tweet like an upper-case percentage in the tweet or time of tweet posted, etc, and user-based features, such as the number of follower/following of the user, etc. From the Tweepy API, we have already extracted past tweets of a user from a list of 76 pre-labeled spammers and non-spammers.

Tweet based features and user-based features together comprise 21 features that are tweet, tweet time, number of followers, number of tweets, number of favorites, number of friends, number of lists, number of screen names used, upper-case percentage use, number of link/URLs used, link to word percentage, same screen name percentage use, tweet similarity percentage, user re-tweet percentage, user average screen name usage, user tweet frequency, user previously re-tweeted or not, user upper-case percentage, user average link/URLs count, user link to word percentage, user link use frequency, user same screen name percentage use, user Tweet length standard deviation, user tweet similarity.

\subsection{Tweet based features}
From only the text of the tweet, we can get a lot of properties like  upper-case percentage, number of screen names in a tweet, link to word percentage, same screen name percentage and tweet similarity. These features correspond to the textual attributes of the tweet which are useful for spam classification.\\
The Tweet similarity percentage \textit{T\begin{scriptsize}s\end{scriptsize}} is calculated as shown in equation (\ref{eq:3}) ,
\begin{equation} \label{eq:3}
\textit{T\begin{scriptsize}s\end{scriptsize}} = \frac{\Sigma \textit{PS\begin{scriptsize}i\end{scriptsize}}}{\textit{n}}
\end{equation}
\begin{equation} \label{eq:4}
\textit{PS\begin{scriptsize}i\end{scriptsize}} = \frac{\textit{n\begin{scriptsize}s\end{scriptsize}}}{\textit{n\begin{scriptsize}t\end{scriptsize}}}\times 100
\end{equation}
Let the tweet of which we are finding the tweet similarity percentage be \textit{T}. We compare \textit{T} having \textit{n\begin{scriptsize}t\end{scriptsize}} words with all the previous  \textit{$i^{\textit{th}}$} tweets \textit{T\begin{scriptsize}i\end{scriptsize}}'s of that user and each comparison is given a percentage of similarity \textit{PS\begin{scriptsize}i\end{scriptsize}} which is percentage of number of similar words \textit{n\begin{scriptsize}s\end{scriptsize}} which are there in both the tweets except all the hashtags, screen names and links to \textit{n\begin{scriptsize}t\end{scriptsize}} and the average of all these \textit{PS\begin{scriptsize}i\end{scriptsize}}'s is calculated to give the tweet similarity percentage of the tweet \textit{T}. This gives how much this tweet is similar to the previous tweets of the user that has posted this tweet.
\begin{figure}[ht]
\begin{center}
\centerline{\includegraphics[width=0.5\textwidth]{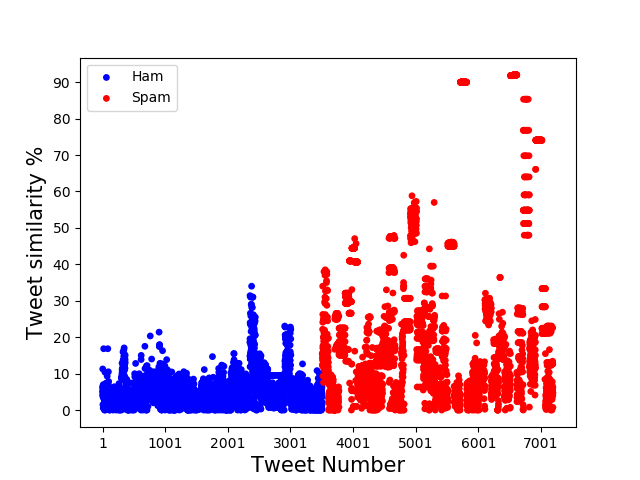}}
\caption{Tweet similarity \% - tweet}
\label{TweetSimilarityPer-tweet}
\end{center}
\vskip -0.2in
\end{figure}
From Fig.\ref{TweetSimilarityPer-tweet} it can easily be seen that spammers post mostly similar tweets but the non-spammers do not. From the dataset it can be seen that spammers only post similar tweets with same agenda every time i.e. to promote a link or user or advertise some of their product.

It is observed from our dataset that most of the spam tweets have a high percentage use of upper-case characters, for example tweets like "\textit{PSP DOWNLOAD Music and Movies HERE! Have the BEST Experience EVER!  http://bit.ly/5QD3Hv}". Upper-case in tweets is usually used to emphasise some part of the tweet. Upper-case percentage \textit{U} can be found by equation  (\ref{eq:5}) where \textit{n\begin{scriptsize}u\end{scriptsize}} is number of upper-case words in the tweet and \textit{n\begin{scriptsize}t\end{scriptsize}} is total number of words in the tweet.
\begin{equation} \label{eq:5}
\textit{U} = \frac{\textit{n\begin{scriptsize}u\end{scriptsize}}}{\textit{n\begin{scriptsize}t\end{scriptsize}}}\times 100
\end{equation}
\begin{figure}[ht]
\begin{center}
\centerline{\includegraphics[width=.5\textwidth]{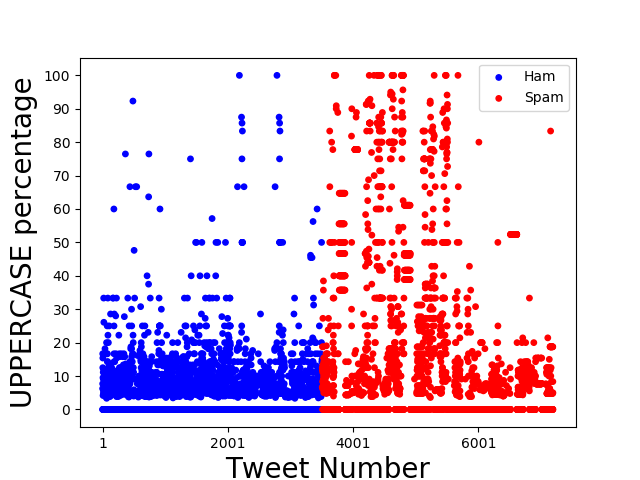}}
\caption{Uppercase \% - Tweet}
\label{UppercasePer-tweet(New)}
\end{center}
\vskip -0.2in
\end{figure}
From Fig.\ref{UppercasePer-tweet(New)} we can easily identify that most of the spammers use high percentage of upper-case words in there tweets but non-spammers don't.

There are some spammers who have a tendency to act as follow train. Follow trains are those types of Twitter accounts which advertise different user accounts by posting a lot of similar kind of tweets which includes requests to follow random users who have started following the follow train. There strategy is to attract different users to follow them by giving them the attractions of gaining followers and then they advertise their products. So number of screen names in a tweet plays an important role for classification of these types of spammers.
\begin{figure}[ht]
\begin{center}
\centerline{\includegraphics[width=.5\textwidth]{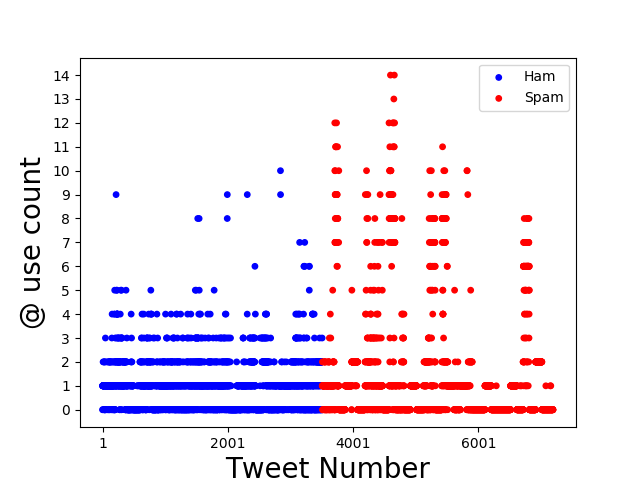}}
\caption{Screen name use count - tweet}
\label{@UseCount-tweet(New)}
\end{center}
\vskip -0.2in
\end{figure}
From Fig.\ref{@UseCount-tweet(New)} we can directly see that most of the tweets from the users use high number of screen names and the reason observed from the dataset is that they have a high tendency of to promote other users but this is not the case with non-spammers.

\begin{figure}[ht]
\begin{center}
\centerline{\includegraphics[width=.5\textwidth]{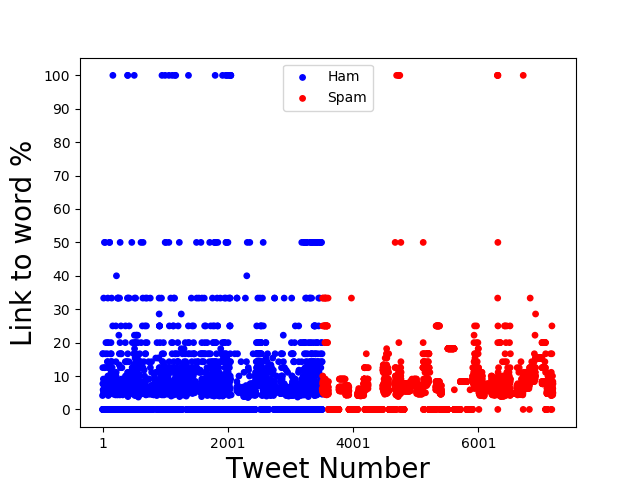}}
\caption{Link to word \% - Tweet}
\label{Link2WordPer-tweet(New)}
\end{center}
\vskip -0.2in
\end{figure}
There are some spammers which have a natural tendency to advertise themselves by providing links to their products in the tweets. Link to word percentage \textit{L\begin{small}2\end{small}W} can be found by equation  (\ref{eq:6}) where \textit{n\begin{scriptsize}l\end{scriptsize}} is number of links in the tweet and \textit{n\begin{scriptsize}t\end{scriptsize}} is total number of words in the tweet.
\begin{equation} \label{eq:6}
\textit{L\begin{small}2\end{small}W} = \frac{\textit{n\begin{scriptsize}l\end{scriptsize}}}{\textit{n\begin{scriptsize}t\end{scriptsize}}}\times 100
\end{equation}

It is observed from Fig.\ref{Link2WordPer-tweet(New)} that spammers usually have low Link to word percentage and the reason for this which is observed from the dataset is that most spammers post tweets which mostly have less words and more link.
\subsection{User based features}
Text of the tweet alone cannot be used for spam classification. From the dataset which we have made, we can see that users who post spam tweets also have different trends in some of the user based properties like re-tweet percentage, link use percentage, percentage of time user tweeted with same screen name, tweet frequency, upper-case use percentage, standard deviation of tweet length, tweet similarity percentage of the user, number of followers and following, number of tweets, number of lists, number of favourites.

\begin{figure}[ht]
\begin{center}
\centerline{\includegraphics[width=.5\textwidth]{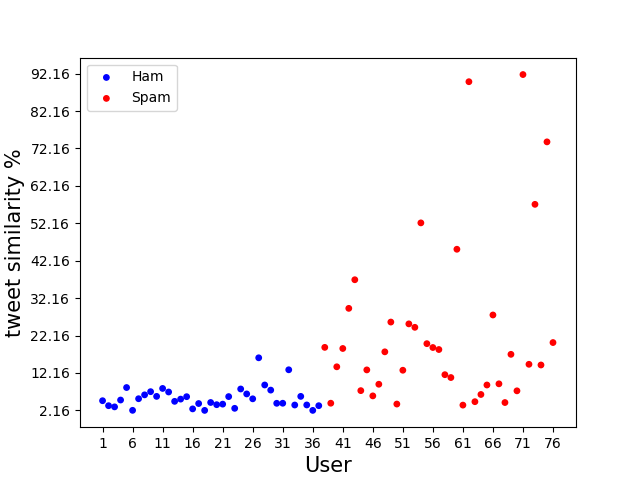}}
\caption{Tweet similarity \% - user}
\label{TweetSimilarity-user}
\end{center}
\vskip -0.2in
\end{figure}

As explained above, tweet similarity percentage plays an important role for spam classification purpose. From Fig.\ref{TweetSimilarity-user} it can easily be seen that spammers post mostly similar tweets but the non-spammers don't. From the dataset it can be seen that spammers only post similar tweets with same agenda every time i.e. to promote a link or user or advertise some of their product. User's tweet similarity percentage can be calculating by taking average of  \textit{T\begin{scriptsize}s\end{scriptsize}}'s of all the tweets of an user where {T\begin{scriptsize}s\end{scriptsize}} is calculated by equation (\ref{eq:3}) \\
A re-tweet is a re-posting of a tweet. Twitter's re-tweet feature helps to quickly share that tweet with all of the followers. One can re-tweet their own tweets or tweets from other users. Sometimes people type "RT" at the beginning of a Tweet to indicate that they are re-posting someone else's content. It is observed from our dataset that most of the non-spam users have a tendency to re-tweet other posts or there own posts, but in the case of spam users they normally do not post re-tweets  however if some of the spammers do re-tweeting then the same tweet is used again and again.

\begin{figure}[ht]
\begin{center}
\centerline{\includegraphics[width=.5\textwidth]{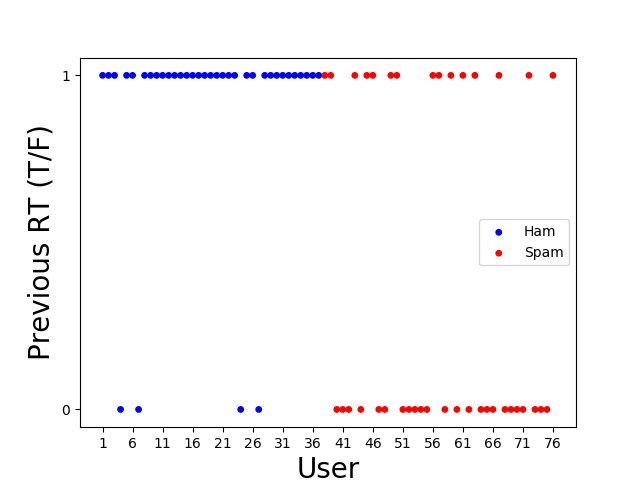}}
\caption{Retweet \% - user}
\label{Retweets-user}
\end{center}
\vskip -0.2in
\end{figure}

\begin{figure}[ht]
\begin{center}
\centerline{\includegraphics[width=.5\textwidth]{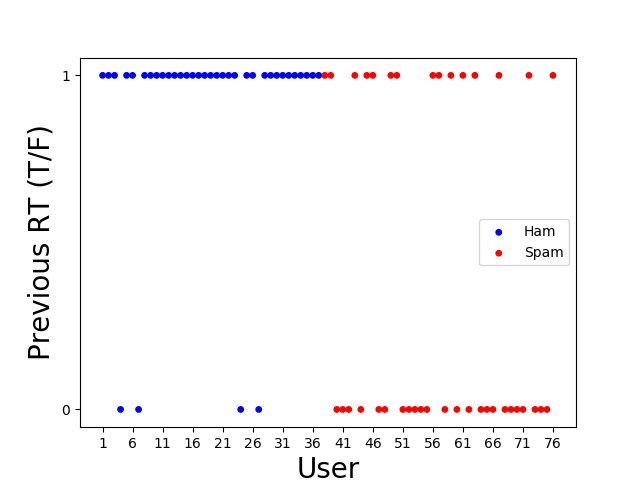}}
\caption{Retweeted before - user}
\label{Retweetedbefore-user}
\end{center}
\vskip -0.2in
\end{figure}

It is observed from  Fig.\ref{Retweets-user} and Fig.\ref{Retweetedbefore-user} that most of the users who are spammers never post re-tweets or if they do then they only post re-tweets of similar other spam accounts only , but on the other hand most of the non-spam users frequently posts re-tweets.\\
\begin{figure}[ht]
\begin{center}
\centerline{\includegraphics[width=.5\textwidth]{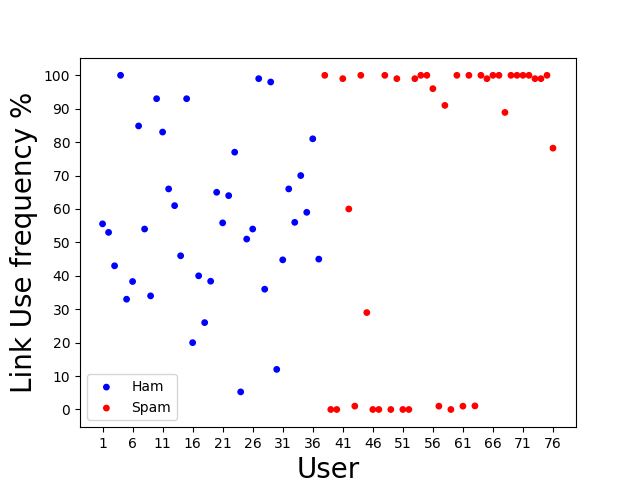}}
\caption{Link use frequency \% - user}
\label{LinkUseFrequency-user(New)}
\end{center}
\vskip -0.2in
\end{figure}

\begin{figure}[ht]
\begin{center}
\centerline{\includegraphics[width=.5\textwidth]{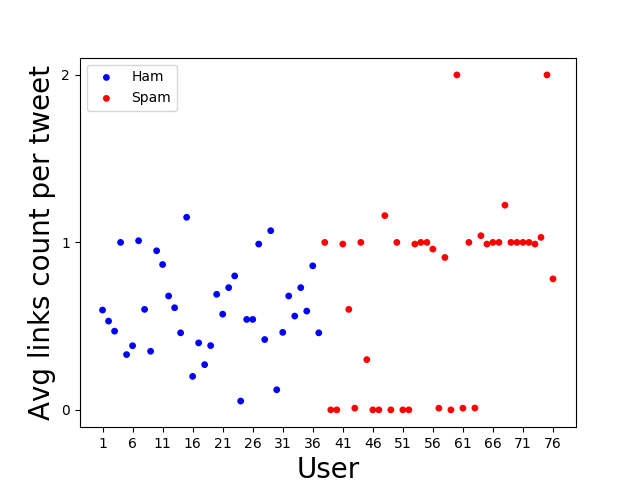}}
\caption{Avg link per tweet - user}
\label{AvgLinkPerTweet-user(New)}
\end{center}
\vskip -0.2in
\end{figure}
Link/URL use frequency percentage \textit{L}, is also an important feature for classification of those type of spammers who usually send a high percentage of tweets having links with high frequency. These type of spammers usually post tweets with similar links which are associated with some product they want to advertise about. It can be calculated by equation  (\ref{eq:7}) where \textit{n\begin{scriptsize}l\end{scriptsize}} is number of previous tweets of the users with any links/URLs and \textit{n\begin{scriptsize}t\end{scriptsize}} is total number of tweets of the users.
\begin{equation} \label{eq:7}
\textit{L} = \frac{\textit{n\begin{scriptsize}l\end{scriptsize}}}{\textit{n\begin{scriptsize}t\end{scriptsize}}}\times 100
\end{equation}
From Fig.\ref{LinkUseFrequency-user(New)} and Fig.\ref{AvgLinkPerTweet-user(New)} we can see that spammers either have very high frequency of tweets with links or very low but that's not the case with non-spammers, It is also observed that those spammers which have a high frequency of link use mostly the same set or same links in each of their post which they want to advertise or if they have a very low link use frequency then they are not advertising any link but it is observed that they mostly advertise other user and most of their posts are related to increasing the follow count of other users.

\begin{figure}[ht]
\begin{center}
\centerline{\includegraphics[width=.5\textwidth]{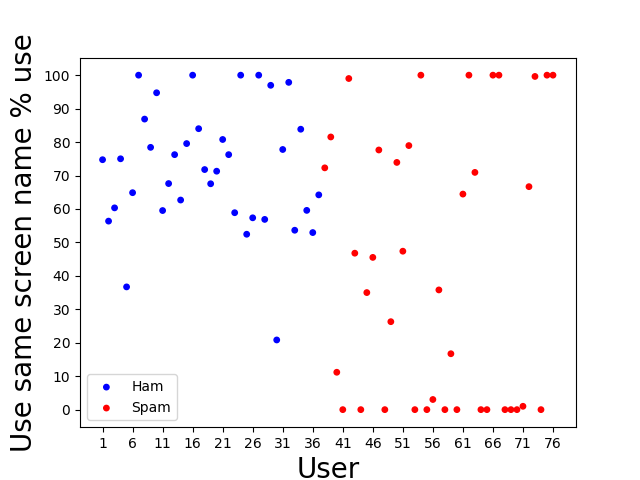}}
\caption{\% time user tweeted with same screen name}
\label{PerUserTweetedSame@Screen-user(New)}
\end{center}
\vskip -0.2in
\end{figure}
Those type of spammers who have a tendency to advertise other users and get there following to advertise their product uses a lot of screen names in there tweets. So percentage of time user tweeted with same screen name \textit{S} is calculated by finding the percentage of number of all the unique screen names \textit{n\begin{scriptsize}us\end{scriptsize}} used to all the number of all screen names \textit{n\begin{scriptsize}ts\end{scriptsize}} used in all the tweets of the user as shown in equation (\ref{eq:8}).
\begin{equation} \label{eq:8}
\textit{S} = \frac{\textit{n\begin{scriptsize}us\end{scriptsize}}}{\textit{n\begin{scriptsize}ts\end{scriptsize}}}\times 100
\end{equation}
From Fig.\ref{PerUserTweetedSame@Screen-user(New)} it is observed that normal users mostly use same screen name in there tweets but spammers either use no screen name, from the dataset it is observed that these are those types of spammers who only promote links in there posts but not advertise users. Other type of spammers are those that always use same screen name in their posts and always want to promote same set of user in there tweets. Rest of the spammer don't use same screen name frequently (They advertise many different users In there tweets).

From Tweepy Twitter API we can also extract the time at which a given tweet was posted. We have used this feature to find tweet frequency of user.

\begin{figure}[ht]
\begin{center}
\centerline{\includegraphics[width=.5\textwidth]{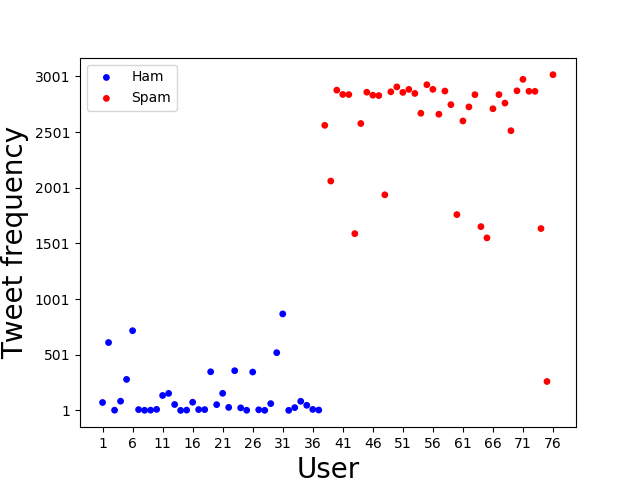}}
\caption{Tweet frequency}
\label{TweetFrequency-user(New)}
\end{center}
\vskip -0.2in
\end{figure}
So from Fig.\ref{TweetFrequency-user(New)} we can easily observe that mostly all the spammers have high tweet frequency of tweet and on the other hand normal users have very low tweets frequency. These types of spammers are usually some type of bots. These types of bots tend to post similar types of tweets.

Tweet length is also an important factor. It is observed from the dataset which we have generated that usually there is a lot of variation in the length of the tweet of the normal user but the length of the tweet of the spammers usually remains the same. So we have also considered the standard deviation of tweet-length as an important user-based feature.

\begin{figure}[ht] 
\begin{center}
\centerline{\includegraphics[width=.5\textwidth]{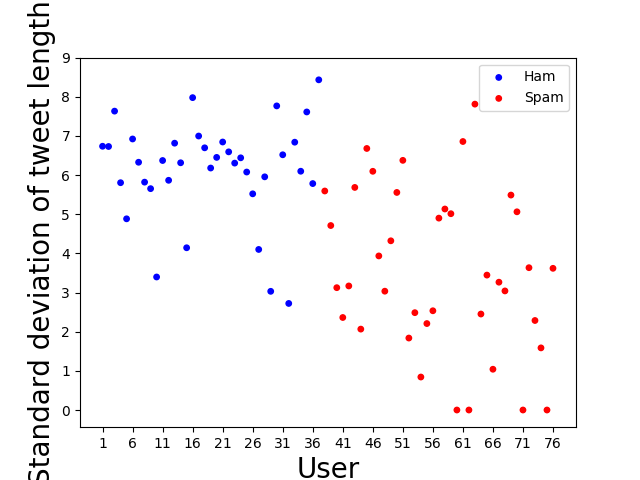}}
\caption{Standard deviation of tweet length}
\label{SDeviationTweetLength-user(New)}
\end{center}
\vskip -0.2in
\end{figure}
From Fig.\ref{SDeviationTweetLength-user(New)} we can see that most of the spammers have a low standard deviation in tweet lengths which signifies that most of there tweets are of the same length and then from observing the dataset we can see that most of them have the same length and format but there is an only minor change like change in the link in the tweet or change in the screen name but this is not the case with normal users.

As we have discussed above that the upper-case use in the tweets is used for emphasizing certain information within the tweets. This is heavily used by spam users to advertise. So we calculate this by finding the percentage of time the user uses upper-case in their tweets.

\begin{figure}[ht]
\begin{center}
\centerline{\includegraphics[width=.5\textwidth]{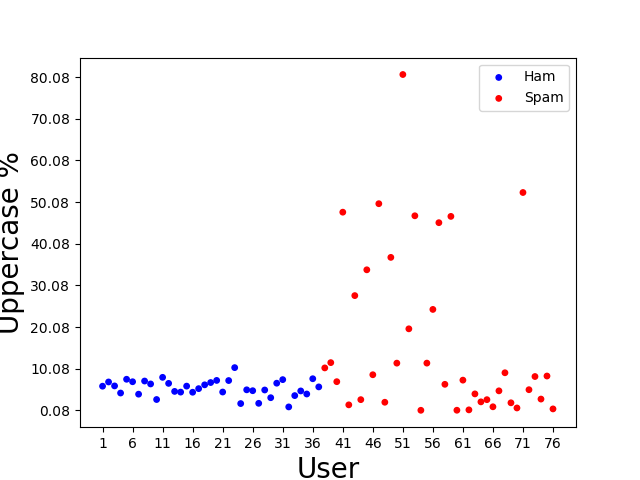}}
\caption{Uppercase \% - user}
\label{UppercasePer-user(New)}
\end{center}
\vskip -0.2in
\end{figure}

From Fig. \ref{UppercasePer-user(New)} we can easily identify that most of the spammers use a high percentage of upper-case words in their tweets but normal users don't. Other properties considered in this paper are tweet similarity percentage, number of followers, and following, number of tweets, number of lists, number of favorites.

\section{Experiment and evaluation}
\begin{figure}[ht]
\centering
\includegraphics[width=0.7\columnwidth]{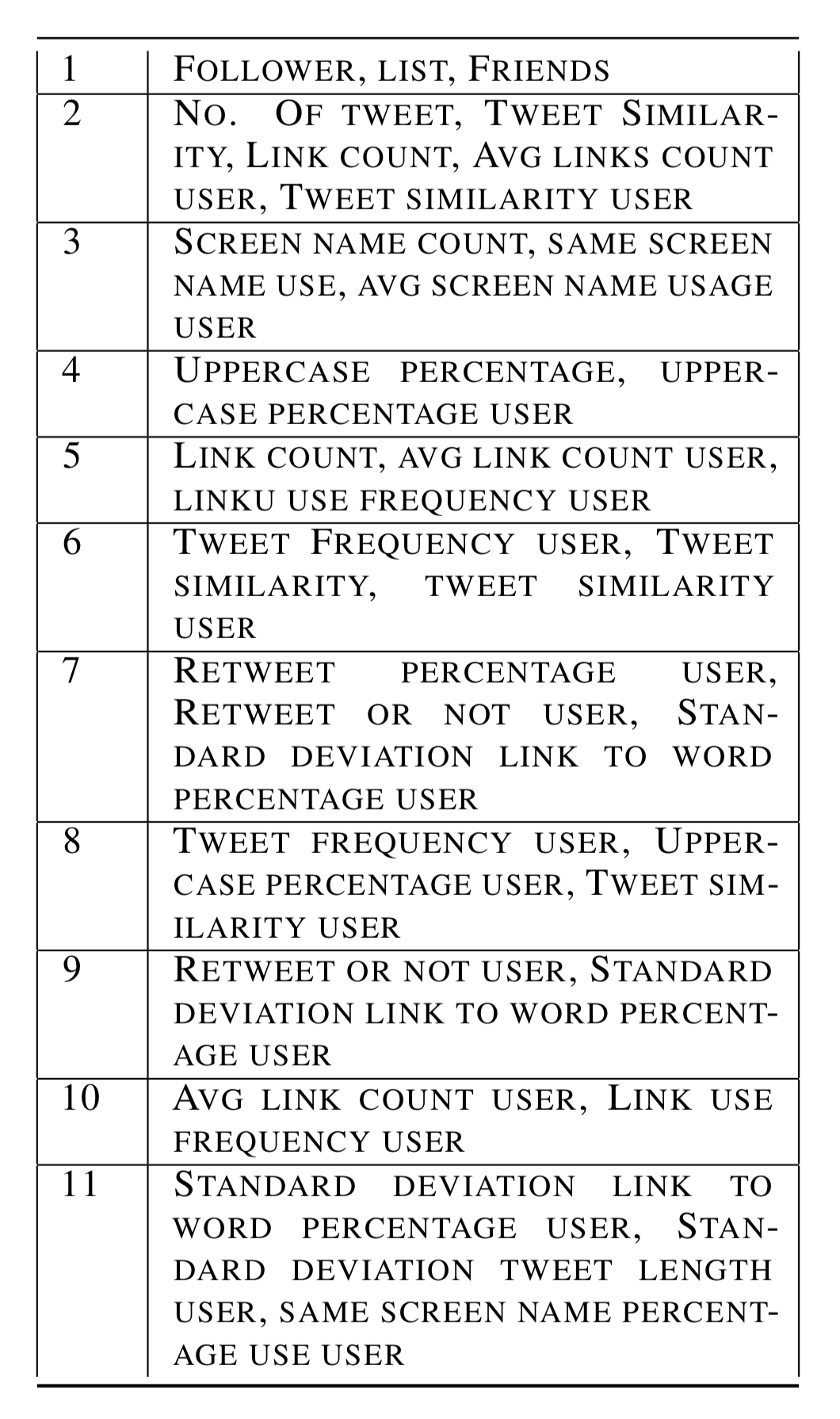}
\caption{Correlated Features}
\label{t1}
\end{figure}
For our experiment we have used the Pearson correlation coefficient for finding the correlation between different features \footnote{https://en.wikipedia.org/wiki/Pearson\_correlation\_coefficient}. In statistics, the Pearson correlation coefficient is a measure of the linear correlation between two variables \textit{X} and \textit{Y}. It has a value between +1 and $-$1, where 1 is a total positive linear correlation, 0 is no linear correlation, and $-$1 is a total negative linear correlation. It is calculated as shown in equation (\ref{eq:pear}).

\begin{equation} \label{eq:pear}
\rho (x,y) =\frac{ cov(x,y)}{\sigma_x \times \sigma_y}
\end{equation}
Here, cov(x,y) is the covariance; $\sigma_x$ is the standard deviation of x, and $\sigma_y$ is the standard deviation of y.

For evaluation of the classifiers we have used 4 metrics that are precision, recall, F1 score and accuracy. Precision and recall are a useful measure of success of prediction when the classes are very imbalanced. In information retrieval, precision is a measure of result relevancy, while recall is a measure of how many truly relevant results are returned.Precision (\textit{P}) is defined as the number of true positives (\textit{T\begin{scriptsize}p\end{scriptsize}}) over the number of true positives plus the number of false positives (\textit{F\begin{scriptsize}p\end{scriptsize}}).
\begin{equation} \label{eq:1}
\textit{P} = \frac{\textit{T\begin{scriptsize}p\end{scriptsize}}}{\textit{F\begin{scriptsize}p\end{scriptsize}} + \textit{T\begin{scriptsize}p\end{scriptsize}}}
\end{equation}
Recall (\textit{R}) is defined as the number of true positives (\textit{T\begin{scriptsize}p\end{scriptsize}}) over the number of true positives plus the number of false negatives (\textit{F\begin{scriptsize}n\end{scriptsize}}).
\begin{equation} \label{eq:2}
\textit{R} = \frac{\textit{T\begin{scriptsize}p\end{scriptsize}}}{\textit{F\begin{scriptsize}p\end{scriptsize}} + \textit{T\begin{scriptsize}n\end{scriptsize}}}
\end{equation}
These quantities are also related to the \textit{F\begin{scriptsize}1\end{scriptsize}} score, which is defined as the harmonic mean of precision and recall. 
\begin{equation}
\textit{\textit{F\begin{scriptsize}1\end{scriptsize}}} = 2\frac{\textit{P}\times\textit{R}}{\textit{P} + \textit{R}}
\end{equation}

\begin{table}[h]
\centering
\resizebox{\columnwidth}{!}{%
\begin{tabular}{l|l|l|l|l}
\thickhline
Model Name   & Precision & Recall & F1 Score  & Accuracy\\
\hline
\hline
\rowcolor{Gray} Our approach & \textbf{98.89}    & 95.57 & \textbf{97.20}    & \textbf{97.57}  \\
SVM               & 91.15    & \textbf{98.84} & 0.9484    & 95.16  \\
Naive Bayes               & 97.05    & 88.67 & 92.67   & 93.70  \\
KNN                & 94.11    & 95.2  & 94.65    & 95.16  \\
Random Forest               & 97.95    & 96.11 & 97.02    & 96.94  \\ 
\thickhline
\end{tabular}}
\caption{Result.}
\label{t2}
\end{table}


From all the above plots we can see that there are a lot of correlated features so to find the correlation between all 21 different features ,we had made a matrix of all the features and found the Pearson's correlation value between different features. From these Pearson's correlation coefficient values we have combined the most correlated features by taking their products. There were 11 sets of correlated features as shown in Table~\ref{t1}. In our model we have used these 11 correlated features and passed it through an artificial neural network having 11 input units, 6 hidden units and one output unit.

\section{Conclusion}
We identified a total of 21 features related to a tweet that contained attributes from both the tweet and the user the corresponding user. While computing the correlation between the features, it was observed that all the features could be grouped into 11 sets of correlated features. Thus our input for the artificial neural network gets reduced to 11 nodes. We apply the ANN for classification on data collected from Twitter API where we used 80\% for training and 20\% for test. This classifier showed better performance than the four other classifiers that we compared with namely SVM, Naive Bayes, K Nearest Neighbours, and Random Forest. On testing, it was observed that precision, F1 score, and accuracy improved for the same dataset with Correlational Artificial Neural Network as shown in Table~\ref{t2}. While we saw a slight decrease in the recall value. However, this can be addressed as a future work where we not only look into features of individual tweets but also look at the links and identify patterns between the tweets and the users. It would be interesting to study if this additional information would improve our results further.

\bibliography{anthology,eacl2021}
\bibliographystyle{acl_natbib}
\end{document}